\documentclass[%
reprint,
amsmath,amssymb,
aps,
longbibliography,
]{revtex4-1}

\pdfoutput=1
\usepackage{amssymb}
\usepackage{mathrsfs}
\usepackage{amsmath,amsfonts,bm,graphicx}
\usepackage{upgreek}
\usepackage{color}
\usepackage{etoolbox}

\usepackage[colorlinks=true,linkcolor=blue,urlcolor=blue,citecolor=blue]{hyperref}

\newcommand{\fig}[1]{Fig.~\ref{#1}}
\newcommand{\Fig}[1]{Figure \ref{#1}}

\newcommand{\del}[1]{{\textcolor{blue}{ [deleted:  #1 ]}}}
\renewcommand{\del}[1]{}

\renewcommand{\section}[1]{}

\newcommand{\ave}{{c}}

\newcommand{\rc}{\mathbf r^{\ave}}
\newcommand{\T}{{T_{traj}}}
\newcommand{\Dr}{{\Delta r}}

\begin{document}
\title{Direct evidence of void induced structural relaxations in colloidal glass formers}

\author{Cho-Tung Yip$^1$}
\author{Masaharu Isobe$^2$}
\author{Chor-Hoi Chan$^1$}
\author{Simiao Ren$^{1,3}$}
\author{Kin-Ping Wong$^3$}
\author{Qingxiao Huo$^1$}
\author{Chun-Sing Lee$^3$}
\author{Yuen-Hong Tsang$^3$}
\author{Yilong Han$^4$}
\author{Chi-Hang Lam$^3$}
\email[Email: ]{C.H.Lam@polyu.edu.hk}
\address{$^1$Department of Physics, Shenzhen Graduate School, Harbin Institute of Technology, Shenzhen 518055, China \\
$^2$Graduate School of Engineering, Nagoya Institute of Technology, Nagoya, 466-8555, Japan\\
$^3$Department of Applied Physics, Hong Kong Polytechnic University, Hung Hom, Hong Kong, China \\
$^4$Department of Physics, Hong Kong University of Science and Technology, Clear Water Bay, Hong Kong, China}

\date{\today}

\begin{abstract}
Particle dynamics in supercooled liquids are often dominated by string-like motions in which lines of particles perform activated hops cooperatively. The structural features triggering these motions, crucial in understanding glassy dynamics, remain highly controversial.
We experimentally study microscopic particle dynamics in colloidal glass formers at high packing fractions. With a small polydispersity  leading to glass-crystal coexistence, a void in the form of a vacancy in the crystal can diffuse reversibly into the glass and further induces string-like motions. In the glass, a void takes the form of a quasi-void consisting of a few neighboring free volumes and is transported by the string-like motions it induces. In fully glassy systems with a large polydispersity, similar quasi-void actions are observed. The mobile particles cluster into string-like or compact geometries, but the compact ones can further be broken down into connected sequences of strings, establishing their general importance.
\end{abstract}

\maketitle


The microscopic origin of kinetic arrest and relaxation mechanisms  in deeply supercooled glassforming liquids have been debated actively for decades  \cite{biroli2013review, stillinger2013review}.
Optical microscopy experiments on colloidal liquids play an important role because   detailed motions of individual particles are accessible \cite{weeks2017review}.
An important progress has been the experimental confirmation of string-like motions
\cite{marcus1999,weeks2000,zhang2011,ganapathy2015}, first discovered in molecular dynamics (MD) simulations \cite{glotzer1998}.
During a string-like event, particles arranged in a line hop to replace the preceding ones.
It can be portraited by a string of participating particles \cite{glotzer1998,glotzer2003,ganapathy2015} or by a smooth line formed by the joint trajectories of these particles \cite{marcus1999,swayamjyoti2014,lam2017}.
A synchronous segment of a string is called a micro-string \cite{glotzer2003}, which has been suggested as elementary relaxations  in glasses \cite{chandler2011,isobe2016,lam2017}. Theoretically, free volumes  \cite{cohen1961} are widely believed to play a pivotal role in glassy dynamics. In particular, voids, i.e. free volumes each of roughly the size of a particle, have long been studied \cite{glarum1960} and are often applied to explain glassy phenomena
\cite{lulli2020spatial}.
However, voids of sizes comparable to the particles show only small correlations to particle dynamics \cite{weitz2005,harrowell2006,swayamjyoti2014}.


We study colloidal liquid sandwiched between two glass plates. Glass-crystal coexisting and glassy systems are obtained using particle size distributions following unimodal (average diameter 3.77$\mu$m) and bimodal (average diameters 3.77$\mu$m and 4.62$\mu$m ) forms respectively.
They are small enough to exhibit strong Brownian motions, but
 are larger than those in typical experiments \cite{marcus1999,weeks2000,zhang2011,ganapathy2015} in order to guarantee a perfect monolayer arrangement of particles upon sinking to the lower plate, forming a quasi-two-dimensional system.
For all results reported here, all particles are practically identified by image analysis software. Their slower movements due to the relatively large particle sizes also enable snapshots taken at rates much faster than the dynamics so that time-averaged particle positions are very accurately measured. We study particle dynamics close to the relaxation time even at a packing fraction, in our knowledge, higher than those used in similar hard sphere experiments previously  \cite{marcus1999,weeks2000,zhang2011,ganapathy2015}. This requires imaging in a single  experiment for up to $10^6$s, compared with typical durations of an hour or less.
See the supplementary information for details \cite{Endnote1}.


\begin{figure*}
\includegraphics[width=0.9\textwidth]{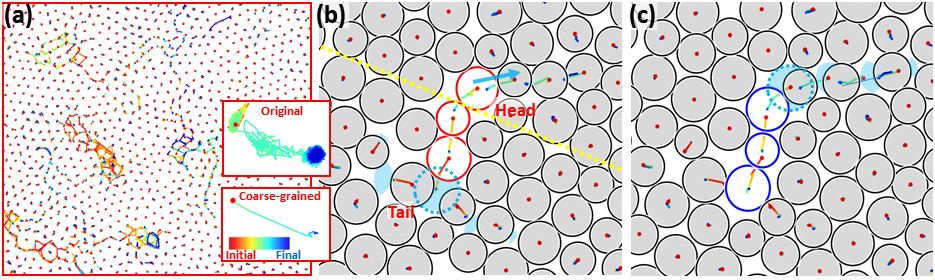}
\caption{(a) Coarse-grained particle trajectories at packing fraction $\phi$  = 0.80 over a duration $\T = 10^6$s.
Line segments in trajectories are colored according to the times of occurrence. Initial positions are denoted by red dots.
Trajectories of hopped particles connect smoothly, revealing string-like motions.
Inset: Magnified views of the original and the coarse-grained trajectories of a typical particle. (b) and (c) Coarse-grained trajectories for duration $\T = 20000$s
showing a typical string-like particle hopping motion. Particle configurations at the beginning (b) and the end (c) of the period are also shown. A quasi-void consisting of fragmented free volumes (blue areas) is transported by the string across a line of otherwise stationary particles (yellow dotted line). Note that the blue dotted circles show the final position of the particle at the string tail in (b) and the initial position of the particle at the string head in (c).
 }
\label{strings}
\end{figure*}

\Fig{strings}(a) shows time-colored  coarse-grained particle trajectories \cite{lam2017} covering a period $\T$ in a glassy colloidal system with bimodal particle sizes at the highest packing fraction $\phi=0.80$. The trajectory of particle $i$ is  based on its coarse-grained position $\rc_i(t)$ at time $t$ obtained by averaging its instantaneous position $\mathbf{r}_i$ over
a coarsening time $\Delta t_\ave$, i.e.
$
 { \rc_i(t) = \langle ~ \mathbf{r}_i(t') ~ \rangle _ {t' \in [t, t+\Delta t_\ave] } }
$.
It is plotted by joining consecutive positions $\rc_i(t)$ and $\rc_i(t+ \Delta t_\ave)$ within time $\T$ with line segments colored according to $t$.
From \fig{strings}(a), a feature characteristic at high $\phi$  is a strong
mobility contrast with  highly mobile particles neighboring nearly stationary ones.
This is verified by MD simulations of hard disks. (See SI for more detailed experimental and simulations results   \cite{Endnote1}.)
We also observe string-like motions \cite{glotzer1998} revealed as smoothly connected trajectories of the mobile particles \cite{marcus1999,swayamjyoti2014,lam2017}. All results presented in the followings are based on coarse-grained particle positions and  trajectories.
\Fig{strings}(b)-(c) magnifies the coarse-grained trajectories showing a typical string-like motion. The initial and final particle configurations of the period are also shown. 
The time-coloring approach conveniently illustrate if the motions are synchronous \cite{lam2017}. For example, the string in \fig{strings}(b)-(c) is asynchronous and consists of three micro-strings \cite{glotzer2003} colored in red, yellow and green.

The high mobility contrast among particles allows easy interpretation of the flow of mass and hence also of free-volume.
Noting that a hopping particle nearly completely replaces the preceding one and adjacent non-hopping particles are practically stationary,
a free volume comparable to the typical particle size is transported by the string through any cross-section of the sample penetrated by the string, e.g.  yellow dotted line in \fig{strings}(b).
For convenience, we designate the direction of a string along the free volume flow direction and that a string points from its tail to head, so that subsequent extension of the string emerges from the head.
Importantly, a continuous free volume of a particle size is not observed, consistent with previous studies \cite{weitz2005,harrowell2006,swayamjyoti2014}. Instead, the transported free volume is fragmented and distributed among a few neighboring interstitial areas both before and after the motion as schematically illustrated in \fig{strings}(b)-(c).
Despite fragmented, the coupled free volumes are transported in whole and behave as a quasi-particle, which we refer to as a quasi-void  \cite{lam2017}. Further explanation of quasi-void is discussed in SI  \cite{Endnote1}.


\begin{figure}
\includegraphics[width=0.48\textwidth]{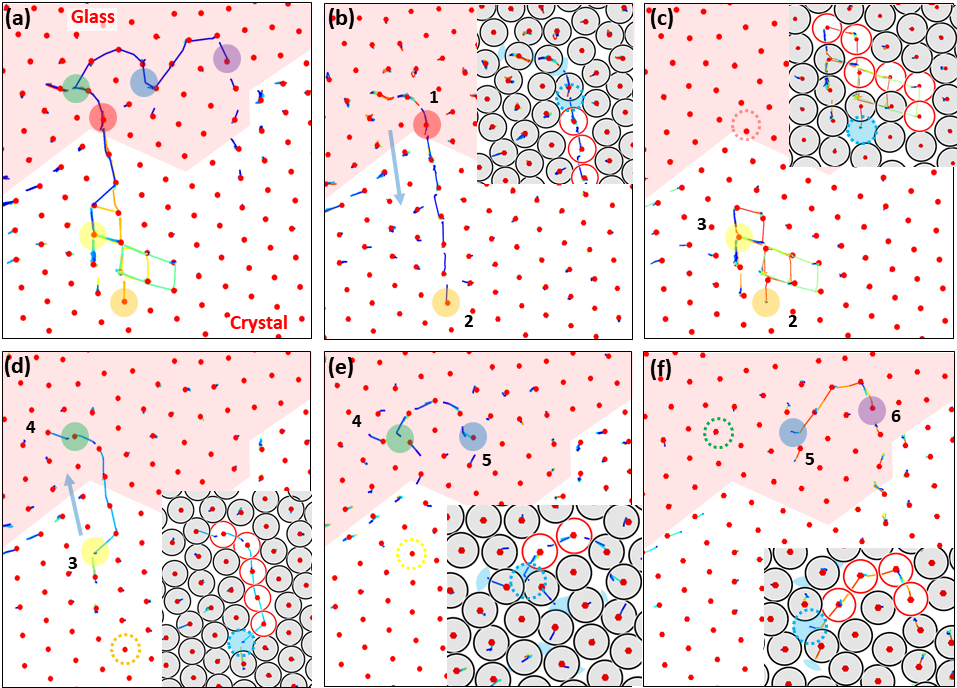}
\caption{
Coarse-grained particle trajectories  covering a time $\T$ = 29120s from a unimodal colloidal system with a small size dispersion exhibiting coexisting glassy (while) and crystalline (red) regions with $\phi$ = 0.81 and 0.83 respectively. (a) A single string-like motion traverses remarkably smoothly across both regions, depicting the motion of a void. (b-f) The same trajectories as in (a) split up over consecutive time sub-intervals showing detailed motions.
Numbered circles mark the initial and final positions of the void at the time subintervals.
 (b) A void moves from the glass to the crystal (blue arrow) and (c) diffuses along  lattice edges. (d) It moves back to the glass (blue arrow) and (e-f) induces a sequence of two micro-string motions. Insets in (b-f): Trajectories shown together with initial particle configurations of the time sub-interval. The void takes the form of a vacancy (blue solid circles) in the crystal (c-d) and a quasi-void with fragmented free volumes (blue areas) in the glass (b,e,f).  
}
\label{crystal}
\end{figure}
The physical relevance of a quasi-void is directly evident from its reversible conversion into a vacancy, an established quasi-particle.
\Fig{crystal}(a) shows particle trajectories in a unimodal colloidal system with  coexisting glassy and crystalline regions. A single string-like motion traverses between the phases. \Fig{crystal}(b)-(f) shows the detailed sequence of events. In the crystal, the trajectories of adjacent hopping particles smoothly connect to form a continuous line, which can be self-crossing and backtracking occasionally. We refer to these joint trajectories in the crystal also as strings, generalizing the usual definition of string-like motions. The vacancy induces such string-like motions during which it is transported from tail to head, analogous to the description above for quasi-voids. Importantly, the string extends remarkably smoothly across the glass-crystal interface.

Remarkably, the excellent continuity of the string even across the phase boundary in \fig{crystal}(a) directly shows that the string-like motion in the glassy region is ultimately caused by the vacancy.
The vacancy upon entering the glass turns into a quasi-void,
which is thus essentially a dressed vacancy.
We refer to both as a void, which manifests itself as a quasi-void or  vacancy in a glass or crystal respectively.


\begin{figure}
\includegraphics[width=0.48\textwidth]{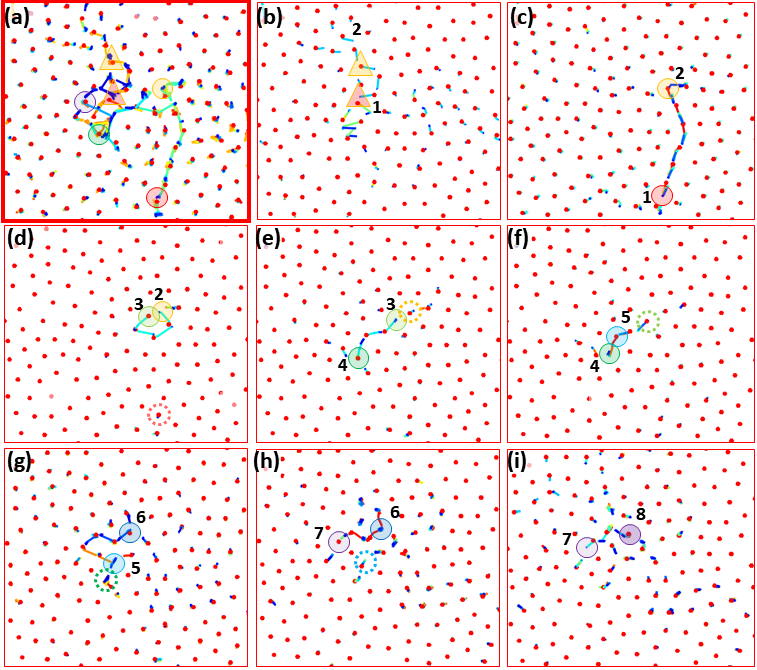}
\caption{
(a) Coarse-grained trajectories in a region from \fig{strings}(a) with a core-like cluster of mobile particles, as revealed by their highly interconnected trajectories.
(b)-(i) Trajectories from (a) split up over consecutive time sub-intervals, showing two independent string-like motions in (b) and (c)-(i).  Numbered triangles and circles mark initial and final positions of the two quasi-voids in each time sub-interval.
}
\label{core}
\end{figure}

String-like motions have been suggested to diminish in importance against compact rearranging groups as supercooling deepens \cite{wolynes2006,ganapathy2015}.
To study it critically, we classify the mobile particles as string-like and core-like and reproduce the trend of an increasing density of core-like particles as observed previously \cite{ganapathy2015} (see SI  \cite{Endnote1}).
\Fig{core}(a) shows the coarse-grained trajectories in a region with a typical core-like group. In sharp contrast to suggestions in  \cite{ganapathy2015},  they can however be broken down into two sequences of string-like motions induced by two quasi-voids as shown in \fig{core}(b)-(k). More examples are shown in the SI.
This fully supports that strings dominate relaxations even at deep supercooling \cite{chandler2011,isobe2016,lam2017}.
The apparent core-like geometries in fact result from an increased dynamic heterogeneity so that strings are more localized and overlap each other. This is also observed in our lattice model of glass characterized by void dynamics \cite{zhang2017}.

\begin{figure}[b]
\centering
\includegraphics[width=3.5in]{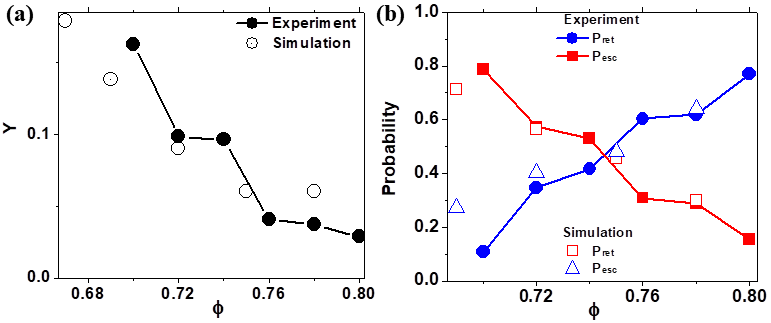}
\caption{(a) Displacement correlation $Y$ measuring the mobility of the neighbors of a hopping particle. At large $\phi$, values tend towards 0, corresponding to existence of stationary neighbors for all hopping particles. (b) Returning and escaping probabilities, $P_{ret}$ and $P_{esc}$, of hopped particles  against $\phi$. They tend towards 1 and 0 respectively at large $\phi$ corresponding to the dominance of back-and-forth hopping motions.
}
\label{corr}
\end{figure}

We now study particle hopping statistics.  A particle is considered hopped if its coarse-grained displacement exceeds $0.8 \sigma$ ($\sigma=3.77 \mu$m), which equals the position of the dip in the van Hove self-correlation function (see SI  \cite{Endnote1}).
At a large packing fraction $\phi$,  dynamics is dominated by particle activated hops organized as strings. Collective flows typical of non-glassy liquids however still exist locally in a diminished role.
To quantify this, we measure a correlation $Y$ of the displacements $\Dr_i$ and $\Dr_j$ of a hopping particle and its neighbors during a time $\delta t$, defined by
\begin{align}
Y = \left< \min \left\{ \Dr_j \right\}_{j \in \Omega_i} / \Dr_{i} \right>_ {i \in \Omega_{hop}} ,
\end{align}
where the minimization is performed over the set $\Omega_i$ of nearest neighbors of particle $i$ while the average is over the set $\Omega_{hop}$ of all hopped particles. \Fig{corr}(a) plots $Y$ against $\phi$. As $\phi$ increases,  $Y$ decreases towards 0, corresponding to the limiting case of particle hops in a background of stationary particles and thus an absence of collective flow. The decline of $Y$ to a small value is verified by our MD simulations (see SI  \cite{Endnote1}). Our result again supports the dominant nature of string-like hopping motions under deep supercooling. It also shows the importance of adopting a  large $\phi$ for suppressing collective flow in order to access true glassy dynamics.

Theoretical descriptions of void induced dynamics traditionally assume free diffusion of voids \cite{glarum1960}. However, we observe strong temporal correlations as revealed by abundant back-and-forth particle hopping motions. They occur not only in our glassy systems but also in the crystalline regions, noting that our unimodal system also admits a small particle size dispersion and hence a non-trivial free energy landscape. Such memory effects is exemplified in the reversed motions in \fig{crystal}(d) versus those in \fig{crystal}(b). To quantify the temporal correlations,
after a particle has hopped, we monitor its subsequent motions for a long time in order to calculate the probabilities $P_{ret}$ and $P_{esc}\simeq 1-P_{ret}$ that it may next perform a returning hop to the original position or an escaping hop to a third position respectively. Results shown in \fig{corr}(b) are in agreement with our MD simulations (see SI   \cite{Endnote1}) and provide an experimental confirmation to   polymer and lattice simulation results \cite{lam2017,zhang2017}. In particular, $P_{ret}$ increases with $\phi$ and reaches a high value of $0.8$ at $\phi=0.80$.  Particle and void dynamics thus differ drastically from free diffusion. A large $P_{ret}$ supports that string-like hopping motions are mostly $\beta$ relaxations and only a vanishingly small fraction of them, i.e. the escaping hops, are  structural relaxations \cite{lam2017,yu2017}.
A theory of glass accounting for these correlations has been suggested recently \cite{lam2018,Deng_2019}.


Strings can extend step by step to long lengths as shown in Figs. (\ref{crystal}) and (\ref{core}).
The extensions are always observed to emerge from the string heads rather than the tails, as is easily explained by the transport of the quasi-void from  tail to head from which further propagations can occur.
Moreover, the long string lengths imply good integrity and long lifetimes of the quasi-voids. This is unexpected because the fragmented free volumes constituting a quasi-void are not energetically bounded and apparently can be easily dispersed. We believe that once a quasi-void, which is usually isolated at a high $\phi$, is momentarily dispersed, the free-volume fragments become immobile. The likely evolution is then the reassembly of the quasi-void, a possibility guaranteed by the time-reversal symmetry of the underlining particle dynamics under local quasi-equilibrium conditions.
The robustness of the quasi-voids distinguishes our description from a generic free-volume picture.
Creation and annihilation of quasi-voids appear to occur mostly in highly active regions with multiple quasi-voids.

The void induced particle dynamics reported here is fundamentally a straightforward generalization of vacancy induced motions in crystals. Vacancies are also believed to be responsible for particle dynamics in high-entropy alloys in crystalline form, which show glass-like sluggish dynamics \cite{yeh2013}. Our work also explains the irrelevance of compact voids  \cite{weitz2005,harrowell2006}, as they have lower entropies and should be statistically unfavorable compared with  quasi-voids.
Dynamics in glasses have been found to be correlated to soft spots and local structures 
\cite{widmer2008,royall2015,PhysRevLett.114.108001,schoenholz2016structural,bapst2020unveiling}. These are consistent with our picture because the presence of a quasi-void clearly impacts the local softness and structures.
The quasi-void notion may shed light on the evolution dynamics of these local properties, which is still lacking.

In conclusion, we have studied quasi-voids in glass formers consisting of fragmented free volumes, which induce and are transported by string-like particle hopping motions. Their relevance is evidenced by the reversible conversion of a quasi-void into a vacancy in a crystalline region. The induced  string-like motions are shown to dominate relaxations even in deep supercooling. We have also presented quantitatively a reduced role of collective flow and drastically enhanced temporal correlations in particle motions as supercooling deepens.


We thank helpful discussions with D.A. Weitz and  J.P. Garrahan. This work was supported by Shenzhen Municipal Science and Technology projects (Grant No. JCYJ 201803063000421), JSPS KAKENHI (Grant No. 17K05574, No. 20K03785), Hong Kong GRF (Grant No. 15330516), and National Natural Science Foundation of China (Grant No. 11974297). Part of the computations was performed at the Supercomputer Center, ISSP, Univ. of Tokyo.\\


%

\end{document}